\documentclass[a4paper,11pt]{article}
\usepackage{pos}
\usepackage{blindtext}
\usepackage{amsmath}
\usepackage{mathtools}
\usepackage{commath}
\usepackage{rviewport}
\usepackage[nottoc]{tocbibind}
\usepackage{lineno}
\usepackage{placeins}
\usepackage[separate-uncertainty=true, table-align-uncertainty=true]{siunitx}
\usepackage{array}
\newcolumntype{P}[1]{>{\centering\arraybackslash}p{#1}}

\usepackage{enumerate}
\usepackage{enumitem}
\usepackage{colortbl}
\definecolor{darkred}{rgb}{0.5,0,0}
\definecolor{darkblue}{rgb}{0,0,0.5}
\definecolor{firebrick}{rgb}{0.75,0.125,0.125}
\definecolor{darkgreen}{rgb}{0,0.5,0}
\usepackage{multirow}
\usepackage{comment}
\usepackage{diagbox}

\newcommand{\til}{{\raise.17ex\hbox{$\scriptstyle\sim$}}}

\newcommand{\NASixtyOne}{NA61\slash SHINE\xspace}

\newcommand{\GeV}{\ensuremath{\text{GeV}}\xspace}

\newcommand{\GeVc}{\ensuremath{\text{GeV}/c}\xspace}

\newcommand{\AGeVc}{\ensuremath{A\,\text{GeV}/c}\xspace}

\newcommand{\pp}{\mbox{\textit{p+p}}\xspace}

\title{First measurements of deuteron production spectra in \pp collisions at beam momentum of 158 GeV/\pmb{c} at \NASixtyOne}
\ShortTitle{First Deuteron Production Spectra in \pp 158~\GeVc at \NASixtyOne}

\manuallySeparateAuthors            % we handle the punctuation ourselves
\author*[a]{Anirvan Shukla}         % [a] → affiliation superscript, * → speaker star
\author{ on behalf of the \NASixtyOne Collaboration}

\affiliation[a]{University of Hawai‘i at Mānoa,\\ Honolulu, Hawai‘i, USA}
\emailAdd{anirvan@hawaii.edu}

\abstract{
The \NASixtyOne spectrometer at the CERN Super Proton Synchrotron (SPS) scans particle production in collisions of nuclei with various sizes at a set of energies covering the SPS energy range towards various physics goals.

This paper presents the first differential production measurements of deuterons at energies relevant for cosmic-ray studies, produced in inelastic \pp interactions at incident projectile momentum of 158~\GeVc ($\sqrt{s}$ = 17.3 GeV). The double-differential spectra are presented as functions of rapidity and transverse momentum and are compared to predictions of the thermal and coalescence models.
These measurements are essential for improving our understanding of cosmic (anti)nuclei production, as detecting cosmic antinuclei can be a breakthrough approach to identifying dark matter. The primary source of cosmic antinuclei background is interactions between cosmic-ray protons and interstellar hydrogen gas. Gaining a deeper insight into the deuteron production mechanism in \pp interactions is an essential first step in modeling cosmic antinuclei production.
}

\ConferenceLogo{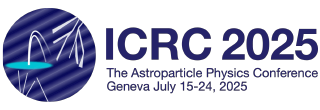}
\FullConference{39th International Cosmic Ray Conference (ICRC2025)\\
 15–24 July 2025\\
Geneva, Switzerland\\}
\begin{document}
\maketitle

%%%%%%%%%%%%%%%%%%%%%%%%%%%%%%%%%%%%%%%%%%%%%%%%%%%%%%%%%%%%%%%%
\section{Introduction}
\label{intro}

Detecting cosmic antinuclei can be a breakthrough approach for identifying dark matter~\cite{doetinchem2020cosmicray}. The primary source of cosmic antinuclei background is interactions between cosmic-ray protons and interstellar hydrogen gas. Gaining a deeper insight into deuteron production in \pp interactions is an essential first step in modeling these astrophysical processes~\cite{Gomez-Coral:2018yuk, Shukla:2020bql}. The two most prevalent formation models, the thermal and coalescence models, are based on different underlying physics. A better understanding of (anti)nuclei production mechanisms is needed, which drives the effort to analyze high-statistics data sets from fixed-target experiments~\cite{Shukla:2859334}.

Unlike antideuterons, cosmic-ray deuterons have been measured~\cite{caprice, Tomassetti_2017, PhysRevLett.132.261001}, and they are the most abundant secondary cosmic-ray species in the Galaxy. Cosmic-ray deuterons disintegrate during nuclear processes in star-forming regions. Therefore, unlike primary cosmic-ray particles like protons and helium, they are not anticipated to be accelerated in supernova remnants. Rather, deuteron formation in cosmic rays (CRs) is understood to be mainly because of CR interactions with the interstellar medium~\cite{gomezcoral2023current, PhysRevLett.132.261001}. Secondary cosmic-ray deuterons are produced chiefly by two processes: fragmentation of CR nuclei such as $^3$He and $^4$He on interstellar hydrogen and helium, and the low-energy resonance reaction
\(p + p \rightarrow d + \pi^{+}\).
%\begin{equation}
%p + p \rightarrow d + \pi^{+}
%\label{eq:resonancedeuteron}
%\end{equation}
Fragmentation is the main source of CR deuterons, because the resonance production channel is only possible at very low collision energies below 1\,GeV~\cite{meyer, Gomez-Coral:2018yuk}.
%Deuteron production in this process happens in a narrow energy distribution (FWHM \til320\,MeV) with a maximum energy of around \til0.6\,GeV~\cite{meyer, Gomez-Coral:2018yuk}.
In addition to these two mechanisms, deuteron production in \pp and $p$+A interactions has also been measured in ground-based collider experiments, for example via:
\begin{equation}
p + p \rightarrow p + \overline{n} + d.
\label{eq:primarydeuteron}
\end{equation}
Of the three deuteron-producing mechanisms described above, the third mechanism (Eq.~\ref{eq:primarydeuteron}) is the only process that also allows the formation of secondary cosmic-ray antinuclei such as antideuterons and antihelium.
(Anti)nuclei production via this mechanism can be modeled using the coalescence model, in which free nucleons resulting from CR--ISM interactions may coalesce to form deuterons~\cite{Gomez-Coral:2018yuk, Shukla:2020bql}.
%Such production sources have not been incorporated yet in the standard calculation of the secondary CR deuteron flux.
%In collider experiments, both resonance and coalescence production can contribute at low momentum~\cite{Gomez-Coral:2018yuk}.
It can also be modeled with the thermal model~\cite{1997hep.ph....2274B, andronic, cleymans, Bellini:2018epz},
%In this model, all products of the fireball continue interacting with each other until the mean free path for elastic collisions is larger than the system size (freeze-out)~\cite{Bellini:2018epz}.
as well as transport models such as AMPT~\cite{Lin_2005, Liu:2022vbg, Shao_2022}. %which have successfully described (anti)nuclei yields.
%initially developed for heavy-ion collisions, can also successfully describe antinuclei production in \pp interactions at RHIC and ALICE energies~\cite{Liu:2022vbg, Shao_2022}.
Examining particle yield measurements in \pp interactions over a large phase space range enables discrimination between different models.
However, existing \pp data in the momentum range relevant to cosmic antideuterons are limited and have significant uncertainties~\cite{Gomez-Coral:2018yuk, doetinchem2020cosmicray}.
%However, this is severely limited by the availability of \pp data and their significant uncertainties in the momentum range relevant to cosmic antideuterons.

%It also has to be noted that measurements in heavy-ion collisions or at high energies~\cite{Alper:1973my,  Henning:1977mt,  SimonGillo:1995dh,  Armstrong:2000gd, Afanasev:2000ku, Anticic:2004yj, Adler:2004uy, Alexopoulos:2000jk, Aktas:2004pq, Asner:2006pw, Schael2006, Abelev:2010, Agakishiev:2011ib, ALICE:2015wav, Adam:2015yta, Adam:2015pna, Acharya:2017dmc, ALICE:2019dgz, Adam:2019phl, Adam:2019wnb, Kittiratpattana_2020, Acharya_2022} are less applicable to evaluate cosmic antideuteron candidates because these interactions are subdominant in the Galaxy~\cite{doetinchem2020cosmicray}.

%The production of (anti)nuclei in \pp collisions can be discussed in a thermal model approach, with the hadronization happening in so-called fireballs~\cite{1997hep.ph....2274B,andronic,cleymans}. All products of the fireball continue interacting with each other until the mean free path for elastic collisions is larger than the system size (freeze-out)~\cite{Bellini:2018epz}. Furthermore, the multi-phase transport AMPT model~\cite{Lin_2005}, initially developed for heavy-ion collisions, was successfully applied to describe antinuclei production in \pp at RHIC and ALICE energies and the production threshold~\cite{Liu:2022vbg, Shao_2022}. In general, examining particle yield measurements in \pp interactions over a large phase space range enables discriminating between different model predictions.

Understanding deuteron production in \pp interactions at energies between 100--400~\GeVc is a necessary first step to understand how CR antinuclei are produced~\cite{Gomez-Coral:2018yuk}. This will also help towards developing a quantum mechanical description of the (anti)nuclei formation process~\cite{Scheibl:1998tk, Blum:2017qnn, Kachelriess:2019taq, Kachelrie__2021, vonDoetinchem:2914265}.
Apart from the importance of these new deuteron measurements for both nuclear physics and in the interpretation of cosmic-ray measurements, this analysis will also serve as the basis of future antideuteron measurements with ultra-high statistics at \NASixtyOne.

\NASixtyOne at the CERN Super Proton Synchrotron (SPS) performs systematic studies of hadron production in strong interactions to address open questions in cosmic-ray physics and in the exploration of the phase diagram of strongly interacting matter. Previous results~\cite{Aduszkiewicz:2017sei} have been used in several cosmic-ray studies, highlighting their importance for interpreting cosmic-ray data~\cite{Korsmeier:2018gcy, Gomez-Coral:2018yuk, Cuoco:2019kuu, Kachelriess:2019ifk, Boudaud:2019efq, Shukla:2020bql, doetinchem2020cosmicray}.

The NA49 experiment~\cite{NA49:1999myq}, the predecessor of \NASixtyOne at CERN SPS, analyzed Pb+Pb interactions at 158\,GeV/$c$ beam momentum. It was demonstrated that both deuterons and antideuterons can be identified in this experimental setup~\cite{dbarna49}.
While measurements from heavy-ion collisions cannot be directly applied to \pp interactions due to the large difference in the system size~\cite{Shukla:2020bql}, the NA49 analysis demonstrated the feasibility of a similar analysis of the high-statistics \pp data sets in \NASixtyOne, which has recorded more than 60 million \pp collisions at beam momentum of 158~\GeVc.

It is worth noting that deuteron production is quite rare in \pp interactions at SPS energies of \til100--400\,GeV. For beam momentum $p_{\text{beam}} = 158$~\GeVc, the coalescence parametrization developed in Refs.~\cite{Gomez-Coral:2018yuk, Shukla:2020bql} predicted a per-event production probability of 0.0004, with an uncertainty band from 0.0002 to 0.0009. The total deuteron production cross section at kinetic energies around \til100\,GeV was estimated in Ref.~\cite{Gomez-Coral:2018yuk} to be close to 0.02\,mb.
%Using these estimates, about \til10,000--60,000 deuterons should have been produced in 60 million \pp interaction events.
To estimate the phase space acceptance for deuterons, about 2.7 trillion \pp interactions were simulated at $p_{\text{beam}} = 158$~\GeVc. Deuteron production in these interactions was simulated by applying the coalescence afterburner condition for seven different values of the coalescence momentum $p_0$~\cite{Shukla:2020bql, Gomez-Coral:2018yuk}.

%The deuteron phase space accessible in \NASixtyOne was found to coincide with peak deuteron production in the center-of-mass frame's backward hemisphere, nevertheless, the detector's efficiency and the narrow phase space acceptance limit the fraction of deuterons that can be tracked and identified to roughly \(1\)--\(2\,\%\) of the total deuterons produced.

The phase space accessible to deuterons coincides with peak deuteron production in the center-of-mass frame's backward hemisphere, but detector efficiency and acceptance limit identification to roughly \(1\)--\(2\,\%\) of the total deuterons produced.

%Hence the full data set was expected to yield \(\mathcal{O}(10^{2})\) reconstructed deuterons.

%However, because of the limited phase space acceptance of the detector, only a fraction of the total deuterons produced can be tracked and identified~\cite{Shukla:2859334, Kowalski:2916893}.

%%%%%%%%%%%%%%%%%%%%%%%%%%%%%%%%%%%%%%%%%%%%%%%%%%%%%%%%%%%%%%%%
\section{Experimental Setup}
\label{sec:exp}

\begin{figure}
\centering
\includegraphics[width=0.8\textwidth]{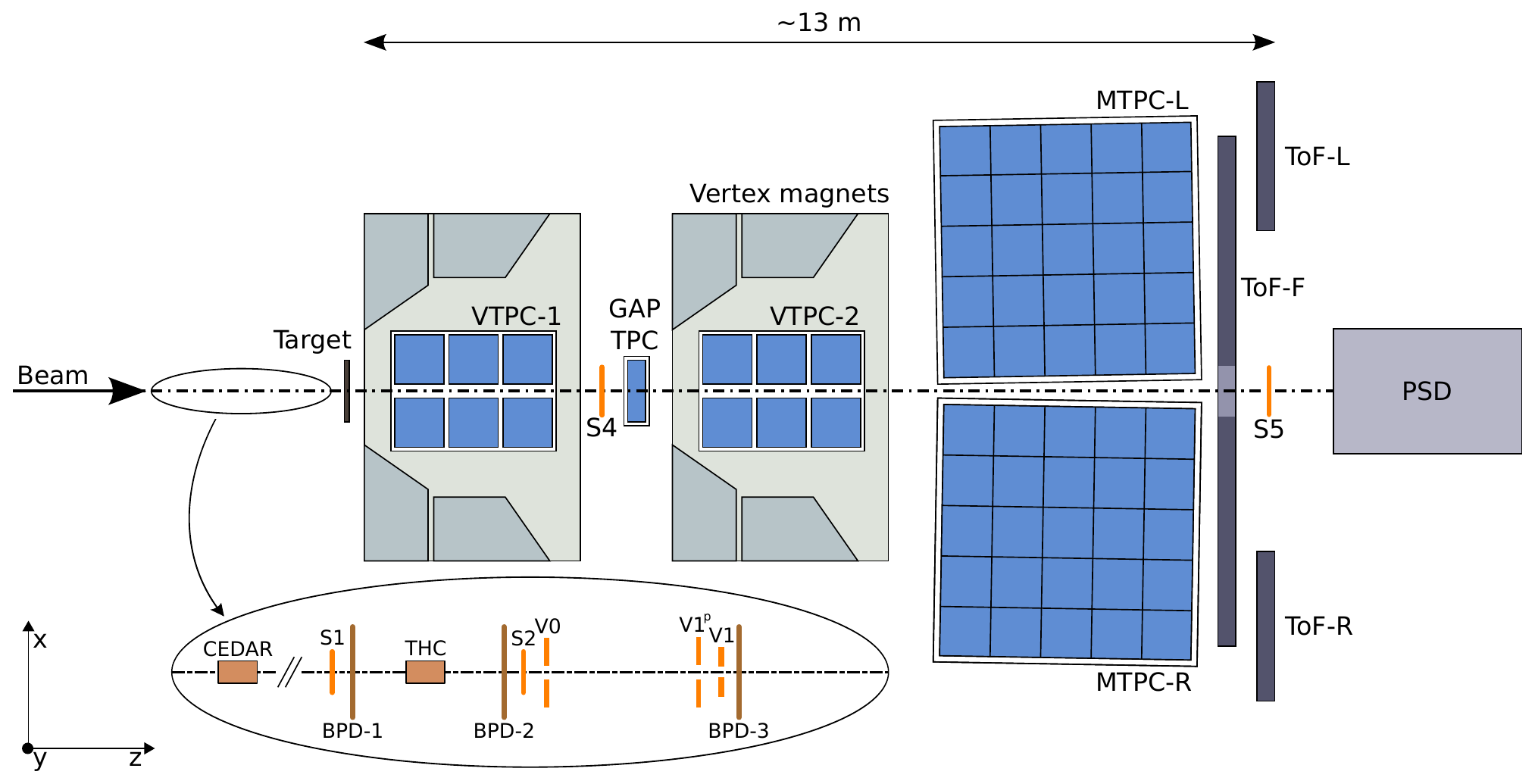}
\caption
    {
    Schematic layout of the \NASixtyOne facility at the CERN SPS used for \pp data taking~\cite{Abgrall:2014fa} (horizontal cut, not to scale).
    The beam instrumentation is sketched in the inset.
    %Alignment of the \NASixtyOne coordinate system is shown in the figure.
    The nominal beam direction is along the $z$ axis.
    The magnetic field bends charged particle trajectories in the $x$-$z$ plane.
    The electron drift direction in the TPCs is along  the $y$ (vertical) axis. 
    }
\label{fig:detector-setup}
\end{figure}

\NASixtyOne receives proton and ion beams from CERN SPS H2 beam line up to 158 \AGeVc, for a broad physics program~\cite{Abgrall:2014fa,Adhikary:2826863}. Dipole and quadrupole magnets guide the beam, and its position is monitored by scintillators S1/S2 and Beam Position Detectors (BPDs).
Particle tracking and identification are provided by three Time Projection Chambers (TPCs): two Vertex TPCs (VTPC-1/2) inside superconducting magnets providing 9\,Tm of bending power, a small GAP TPC, and two large Main TPCs (MTPC-L/R).
%The TPCs run with an argon–carbon dioxide gas mixture and enable 3D tracking  with good momentum resolution.
These are augmented by two downstream time-of-flight (ToF) detectors, ToF-L and ToF-R.

%Upstream of the target, two scintillator beam counters (S1 and S2) measure the incoming beam. The S4 counter is located downstream of the target on the beamline, after VTPC-1. To ensure that the recorded events have a main interaction vertex within the liquid hydrogen target (LHT), a minimum bias trigger is used. This is done by requiring the detection of an incident beam proton upstream of the LH target (measured via signal from the S1 counter) and its disappearance downstream of the target (when there is no signal from the S4 counter). The absence of the beam particle signal in S4 is used to select interactions of beam particles in the target. Together, these define the trigger logic and ensure the rejection of elastically scattered beam particles. The detailed detector setup, trigger logic, and complete operational aspects for this data set are described in Refs.~\cite{Aduszkiewicz:2017sei, Shukla:2859334}.

A minimum-bias trigger selects interactions in the liquid-hydrogen target (LHT) by requiring an incident beam proton in S1 and its disappearance downstream of the target in the S4 counter, suppressing elastic events~\cite{Aduszkiewicz:2017sei}. Detailed detector and trigger descriptions are provided in Refs.~\cite{Abgrall:2014fa, Aduszkiewicz:2017sei, Shukla:2859334}.

%%%%%%%%%%%%%%%%%%%%%%%%%%%%%%%%%%%%%%%%%%%%%%%%%%%%%%%%%%%%%%%%
\section{Data Selection}
\label{sec:data}
This deuteron analysis is based on the \pp data sets collected at \NASixtyOne in 2009, 2010, and 2011. A total of more than 60 million \pp collisions at beam momentum of 158~\GeVc were recorded, which yielded more than 750 million reconstructed particle tracks.
%Deuteron production is rare in \pp interactions at SPS energies of \til100--400 \GeV. For $p_{\text{beam}}$ = 158~\GeVc, the coalescence parametrization developed in Refs.~\cite{Gomez-Coral:2018yuk, Shukla:2020bql} predicted a per-event production probability of 0.0004, with an uncertainty band from 0.0002--0.0009. Using these estimates, about \til 10,000--60,000 deuterons should have been produced in the high-statistics \pp data. However, because of the limited phase space acceptance of the detector, only a fraction of the total deuterons produced can be identified.

The overall analysis procedure consists of the following steps:
\begin{enumerate}[itemsep=1pt, topsep=2pt, parsep=2pt] % local override
\item Event and track selection.
\item Calculation of raw (i.e., uncorrected for biases) particle count distributions of identified charged hadrons.
\item Applying data-derived and simulation-based corrections to the raw counts.
\item Calculation of the corrected particle spectra.
\item Estimation of statistical and systematic uncertainties.
\end{enumerate}

Events with inelastic \pp interactions were selected based on five key criteria. First, the presence of off-time beam particles was rejected by identifying multiple beam signals within $\pm2\,\mu\text{s}$ around the primary trigger particle. Second, the beam proton trajectory was validated by ensuring sufficient signals in the Beam Position Detectors (BPDs), particularly BPD3 which is nearest to the target. Third, each event was required to contain at least one track originating from the reconstructed main \pp interaction vertex. Fourth, the $z$-position of the main vertex was constrained to lie within $\pm20$\,cm of the target center. Finally, events were rejected if they contained only one positively charged track with momentum matching that of the incident beam, thus removing elastic scattering candidates.
%In the Monte Carlo (MC) analysis, the same selection cuts were applied, except for those relating to the Wave Form Analyzer (WFA) and BPD signals, which were not simulated. An effective T2 trigger cut was imposed in the MC by requiring no deposited energy in the simulated S4 counter.
These combined selection steps ensured a consistent dataset of purely inelastic \pp interactions for both data and simulation analyses.

\begin{table}[t]
\centering
\caption{Summary of the recorded high-statistics \pp interactions at $p_{\text{beam}}$ = 158\,GeV/$c$ in \NASixtyOne.}
\small
\begin{tabular}{ |P{1.0cm}|P{5.0cm}|P{5.0cm}|}
 \hline
 Year  & \pp interactions recorded & Total selected events\\
 \hline
 2009  &                    3.8 million &           1.6 million\\
 \hline
 2010  &                   43.1 million &          25.6 million\\
 \hline
 2011  &                   13.7 million &           8.1 million\\
 \hline \hline
 Total &                   60.6 million &          35.3 million\\
 \hline
\end{tabular}
\label{table:ppdatastats}
\end{table}

Reconstructed particle tracks were selected to ensure only primary charged hadrons were retained 
and to minimize contamination from off-time interactions, weak decays, and secondary interactions. 
First, each track had to originate from the main interaction vertex, with a convergent momentum fit. 
%Next, the fitted $x$-component of the rigidity $(p_{\mathrm{lab},x}/q)$, where $(p_{\mathrm{lab},x}$ is the total momentum in the laboratory frame and $q$ is the electric charge, was required to be positive, selecting tracks that reduce uncertainties in the track fit and cluster reconstruction~\cite{PPodlaskiThesisRST}.
Next, track were required to satisfy the condition \(p_{\mathrm{lab},x}/q > 0\), where \(p_{\mathrm{lab},x}\) is the \(x\)-component of the total momentum in the laboratory frame and \(q\) is the particle's charge. This positive-rigidity cut preferentially retains tracks with smaller uncertainties in the track fit and cluster reconstruction~\cite{PPodlaskiThesisRST}.
Tracks also needed at least 30 reconstructed clusters, 
including a minimum of 15 clusters in the Vertex TPCs (or at least 4 in the Gap TPC), 
and a similarly stringent requirement for $\text{d}E/\text{d}x$ clusters. 
Finally, the distance between the main vertex and the track extrapolation in the transverse plane 
(impact parameter) had to be within 4\,cm horizontally and 2\,cm vertically. 
These criteria ensured precise momentum measurement and reliable track identification of primary charged hadrons.

%%%%%%%%%%%%%%%%%%%%%%%%%%%%%%%%%%%%%%%%%%%%%%%%%%%%%%%%%%%%%%%%
\section{Data Analysis}
\label{sec:analysis}

After the application of event and track quality cuts, about 35 million collision events containing about 60 million particle tracks were selected for analysis. The recorded event statistics are summarized in Table~\ref{table:ppdatastats}.

%%%%%%%%%%%%%%%%%%%%%%%%%%%%%%%%%
\subsection{Particle Identification Based on Time of Flight and Energy Loss Measurement}
\label{tofdEdxPID}

Deuterons were identified with the $tof-\text{d}E/\text{d}x$ method, which combines the energy loss measurements ($\text{d}E/\text{d}x$) in the TPCs with the $m^2$ measurements from the ToF detectors. A clean deuteron signal was extracted in regions of large negative rapidity, in the backward hemisphere of the center-of-mass frame (approximately $y = -1$).

%The $tof-\text{d}E/\text{d}x$ analysis requires additional track selection cuts which are specific to the ToF systems. 
%This ToF measurement is calibrated by extrapolating particle tracks in the MTPCs to the ToF walls. The extrapolated scintillator pixel's TDC and ADC channels are checked if they are above threshold.
%The ToF measurement is calibrated by extrapolating MTPC tracks to the ToF walls, requiring TDC and ADC signals from the extrapolated scintillator pixel to be above threshold.
%More detailed corrections take into account the charge deposition and relative position of the incident particle in each ToF pixel. These were applied after the standard event and track selection cuts described earlier. 
After the standard event and track selection cuts described in Sec.~\ref{sec:data}, ToF–specific criteria ensured precise track–pixel matching and noise suppression~\cite{Shukla:2859334, Kowalski:2916893}. These additional cuts are:
\begin{enumerate}[itemsep=1pt, topsep=2pt, parsep=2pt] % local override
\item The hit in the ToF scintillator pixel must be associated with only one reconstructed TPC track.
\item The ToF hit pixel must coincide with the extrapolated pixel where the MTPC track, linearly extrapolated from its last ten clusters in the $x$--$z$ and $y$--$z$ planes, intersects the ToF wall.
%The hit pixel must be the same as the extrapolated pixel, which is found by extrapolating the reconstructed MTPC track to the ToF wall. For this technique, the last ten clusters of all MTPC tracks were fitted with straight lines in the $x$--$z$ and $y$--$z$ planes. The fits were extrapolated to the $z$-positions of all ToF pixels to find the extrapolated pixel.
\item The hit pixel must have corresponding ADC and TDC measurements~\cite{PPodlaskiThesisTOFcuts}.
\item The last point measured on the MTPC track was required to be in the last two padrows of the MTPC
%i.e., with $z >$~ 575\,cm,
to ensure good matching between the TPC tracks and ToF walls.
\item Track residual distance was calculated between the reconstructed global particle trajectory and the last point measured on the track in the MTPC, and it was required to be less than 2\,cm.
\item Finally, the minimum overall efficiency of a ToF pixel used in these measurements was set to 50\%. Pixel with hit efficiencies less than 50\% were categorized as dead, and were not used for measurements. 
\end{enumerate}

Applying these conditions provided reliable extrapolation of the MTPC trajectories toward the ToF detectors, and helped in reducing noise in the selected tracks.

%%%%%%%%%%%%%%%%%%%%%%%%%%%%%%%%%
\subsection{Deuteron Identification using \texorpdfstring{\pmb{$m^2$}}{m2} Distributions}
\label{deuteronPID}
%\texorpdfstring{$m^2$}{m2}

The standard particle identification (PID) technique in the $tof-\text{d}E/\text{d}x$ analysis relies on two-dimensional fits performed in the $m^2$--$\text{d}E/\text{d}x$ plane and was successful in identifying charged hadrons in \pp and Ar+Sc data sets~\cite{Aduszkiewicz:2017sei, 2024ArScNA61}. However, this technique is unable to estimate the deuterons under the proton tail.

Therefore, a new data-driven template fitting method was developed for particle identification~\cite{Grebieszkow:2878507}. For each total momentum and transverse momentum ($p_{\text{tot}}$, $p_{\text{T}}$) bin, pions and positrons were separated from the other particles using the $\text{d}E/\text{d}x$ information by applying cuts on the $\text{d}E/\text{d}x$--momentum plane. As the mass distributions for kaons, protons, and deuterons overlap, the pion mass distribution was modified for detector resolution effects and the kaon, proton, and deuteron mass, respectively. These three modified mass templates served as the input for the combined one-dimensional $m^2$ template fit of the higher-mass $Z=1$-particle mass spectra.

\begin{figure}[t]
  \centering
  \begin{tabular}{ll}
    \includegraphics[width=0.47\textwidth]{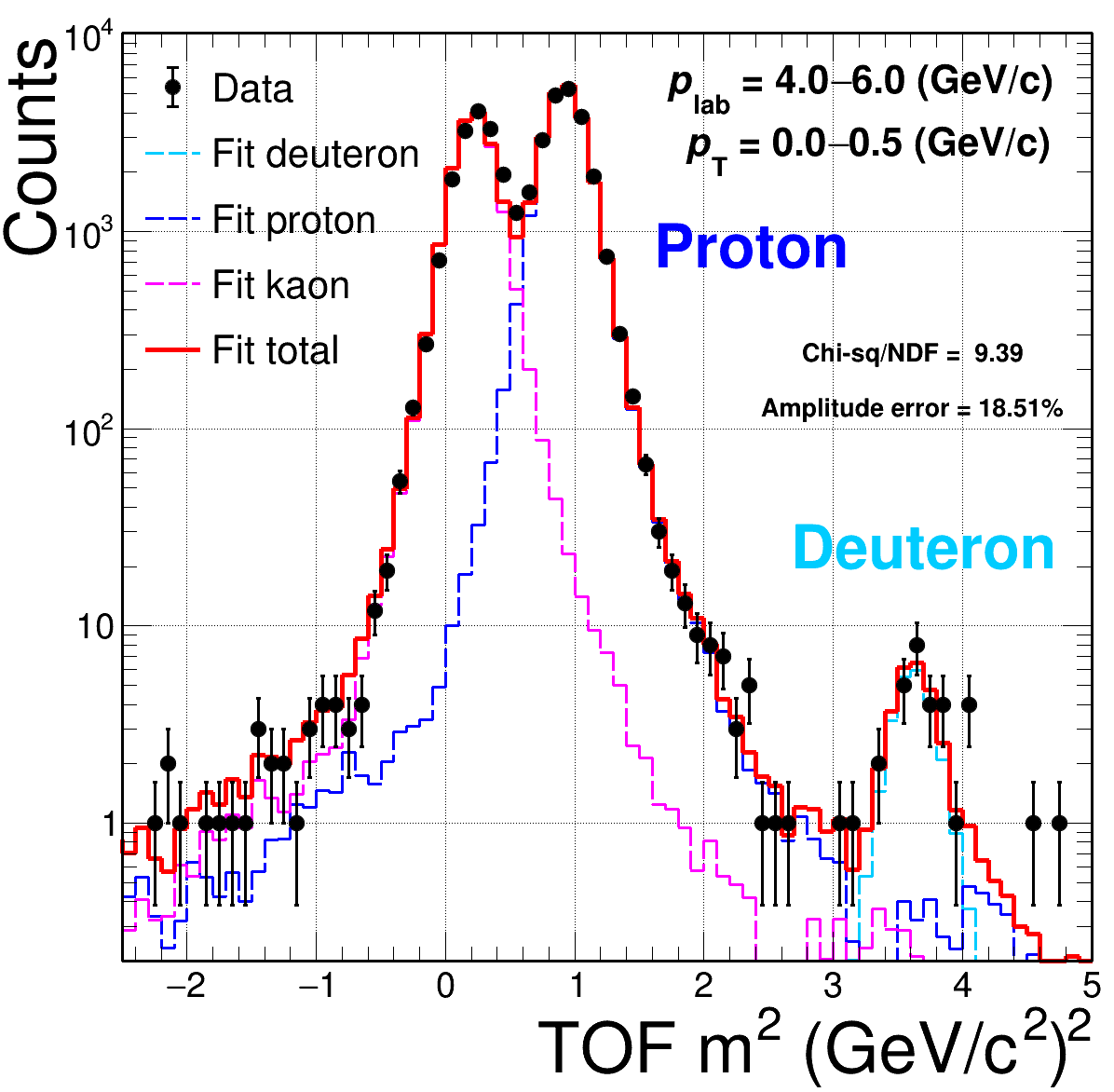} &
    \includegraphics[width=0.47\textwidth]{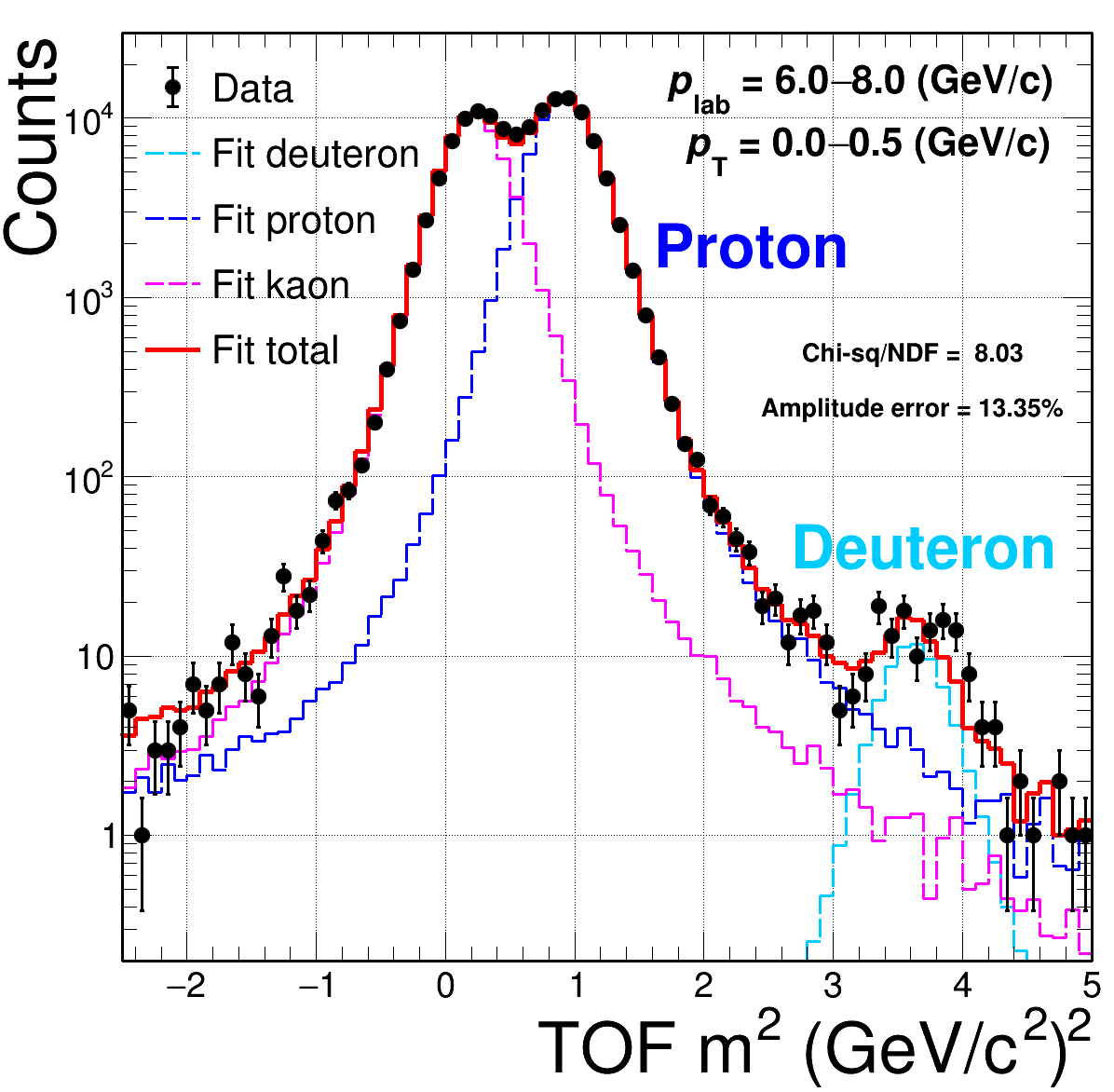}
  \end{tabular}
  \caption[Data-Driven Deuteron Mass Template Fits]{Two example phase space bins with data-driven pion-mass templates to fit the kaon, proton and deuteron peaks.}
  \label{fig:m2_fits}
\end{figure}

Figure~\ref{fig:m2_fits} shows two example ($p_{\text{tot}}$, $p_{\text{T}}$) bins with a clear deuteron peak. It also illustrates the importance of realistically estimating the proton tail in the deuteron mass region. The kaon, proton, and deuteron yield extraction is based on the probability method (Sec. \ref{deuteronProbabilityMethod}).

%\begin{figure}[t]
%  \centering
%  \begin{tabular}{lll} % three columns, not two
%    \includegraphics[width=0.32\textwidth]{figures/Vol-3_massSq_ProtonKaonDeuteron_HighStatistics.png} &
%    \includegraphics[width=0.32\textwidth]{figures/Vol-4_massSq_ProtonKaonDeuteron_HighStatistics.png} &
%    \includegraphics[width=0.32\textwidth]{figures/Vol-5_massSq_ProtonKaonDeuteron_HighStatistics.png}
%  \end{tabular}
%  \caption[Data-Driven Deuteron Mass Template Fits]{Data-driven pion-mass template fits to the kaon, proton and deuteron peaks.  A clear deuteron signal is visible in both example kinematic bins; estimating the proton tail beneath the deuteron peak is critical.}
%  \label{fig:m2_fits}
%\end{figure}

%%%%%%%%%%%%%%%%%%%%%%%%%%%%%%%%%
\subsection{Deuteron Yields with the Probability Method}
\label{deuteronProbabilityMethod}

The estimation of the deuteron yields was done using a modified probability method~\cite{Aduszkiewicz:2017sei, 2024ArScNA61, Shukla:2859334}. In this method, the fits of the $m^2$ distributions were used to compute the probabilities $P_i$ for a track being a given particle type $i$ = $p, K, d$. After obtaining the $m^2$ fits described above, these probabilities can be calculated as:
\begin{equation}
P_{i}(m^2,p_{\text{lab}},p_{\text{T}})=\frac{\rho_{i}(m^2,p_{\text{lab}},p_{\text{T}})} {\sum\limits_{i=p, K, d}^{} \rho_{i}(m^2,p_{\text{lab}},p_{\text{T}})},
\label{eq:prob_tofdedx}
\end{equation}
where $\rho_{i}$ is the value of the fitted function for particle-type $i$ in a given $(p_{\text{lab}},p_{\text{T}})$ bin, evaluated at the $m^2$ of the particle. The denominator represents the total fit value for that bin at the particle's $m^2$. It is assumed that pions and positrons were perfectly identified using the cuts on $\text{d}E/\text{d}x$--momentum plane.
%The probability method was used to assign five probability values to each track in data, for being a pion or a positron, or for being kaon-, proton-, or deuteron-like.
The probability method assigned each track probabilities of being a pion, positron, kaon, proton, or deuteron.

%The probabilities of track identification in the $tof$--\dEdx analysis for positive-charged tracks produced in $\pp 158\,GeV/$c$ are shown in Fig.~\ref{fig:prob_tofdedx}.
The probabilities are either 0 or 1 if the particle was identified as a pion or an electron. Because of the overlapping $m^2$ distributions, the calculated particle probability values for kaons, protons and deuterons range between 0 and 1. Using this method, the fits performed in ($p_{\text{tot}}$, $p_{\text{T}}$) bins were used to get count distributions in $(y,p_{\text{T}})$ bins. The total number of raw (i.e., uncorrected) identified particles $n^{\text{raw}}$ of particle-type $i$ in a given kinematic bin (e.g., $(y,p_{\text{T}})$) is given by~\cite{Rustamov:2012bx}:
\begin{equation}
n^{\text{raw}}_{i=\pi,K,p,d}=\sum_{j=1}^{N_{\text{trk}}}P_{i},
\label{eq:ntrktofdEdx}
\end{equation}
where $P_i$ is the probability of particle type $i$ given by Eq.~\ref{eq:prob_tofdedx}, and $j$ is the index to sum over all $N_{\text{trk}}$ weights in the given kinematic bin.

Approximately \(10\%\) of the statistics were recorded with the liquid-hydrogen target removed to quantify off-target backgrounds. The target-removed sample was scaled by the ratio of events with fitted vertices in the range \(-400 < z < -200~\mathrm{cm}\) between target-inserted and target-removed runs, and its yield was then subtracted for each bin of total momentum and transverse momentum.
After background subtraction, the final particle counts were measured in two-dimensional bins of rapidity ($y$) and transverse momentum ($p_{\text{T}}$).
About 200 deuteron tracks could be identified.
The particle rapidity was calculated in the collision center-of-mass frame. The bin size was chosen taking into account the statistical uncertainties and the resolution of the momentum reconstruction, to reduce the effect of bin-to-bin migration to less than 1\%~\cite{Abgrall:2013pp_pim}.
To finalize the particle spectra, $(y,p_{\text{T}})$-dependent correction factors for detector geometry and ToF-related efficiencies were calculated using Monte Carlo simulations, and applied to correct the normalized raw count distributions.

%%%%%%%%%%%%%%%%%%%%%%%%%%%%%%%%%
\subsection{Estimation of Systematic and Statistical Uncertainties}
\label{deuteronUncertainties}
A preliminary evaluation of both statistical and systematic uncertainties was performed for the preliminary deuteron spectra.

Statistical uncertainties were obtained by treating the raw, background-subtracted total deuteron counts in each $(y,p_{\text{T}})$ bin as independent Poisson variables. The error was subsequently propagated through the spectra calculation. The resulting relative statistical errors are large and range from approximately $30\%$ to $60\%$, reflecting the limited statistics available in these kinematic regions.

The dominant sources of systematic uncertainties are those associated with the one-dimensional $m^2$ fits and the modeling of the proton tail under the deuteron peak.
The deuteron signal sits on a broad background arising primarily from the
high-mass tail of the proton $m^{2}$ distribution.  A data-driven template
fit is employed to model this tail.  The template shape parameters are varied within the detector resolution envelope, and the deuteron yield is re–extracted for each variation. The maximum spread of the refitted yields, about $\pm 20\%$, is assigned as the systematic uncertainty associated with the background proton subtraction.

The baseline analysis employs fixed‐width bins of $\Delta p_{\text{lab}} = 2~\text{GeV}/c$ and $\Delta p_{\text{T}}=0.5~\text{GeV}/c$ to perform the $m^{2}$ fits. Ten alternative grids are formed by halving one or both of these widths.
The root‐mean‐square (RMS) spread of the resulting deuteron yields is $15\%$, which is taken as the systematic uncertainty due to the choice of phase‐space slicing.
%Secondary deuteron production from interactions with upstream detector material was examined using target‐out runs and vertex cuts; its contribution is below~$1\%$ and is therefore neglected.
The two leading systematic contributions are uncorrelated and are therefore combined in quadrature:
\[
  \Delta_{\text{syst}} = \sqrt{(20\%)^{2} + (15\%)^{2}} \approx 25\%.
\]
This bin‐independent envelope is quoted alongside the Poisson statistical bars for every spectra point.

%%%%%%%%%%%%%%%%%%%%%%%%%%%%%%%%%
\subsection{Validation of Methods}
\label{validationOfMethods}
The new techniques in this analysis were validated in two independent ways.
First, in each ($p_{\text{tot}}$, $p_{\text{T}}$) bin, the fitted proton counts from the new fitting technique were compared to the fitted proton counts from the standard method of making two-dimensional fits. This ratio was found to be almost-exactly 1 in almost every ($p_{\text{tot}}$, $p_{\text{T}}$) bin, with a very small associated uncertainty.
Second, the identified proton spectra from the new $tof-\text{d}E/\text{d}x$ technique were compared to the previously-published proton spectra for all available overlapping phase space bins. The comparison showed that the new proton spectra from the $tof-\text{d}E/\text{d}x$ analysis overlapped with published measurements within the uncertainty bands~\cite{Kowalski:2916893}.
Both these checks were used to validate the new technique.

%%%%%%%%%%%%%%%%%%%%%%%%%%%%%%%%%%%%%%%%%%%%%%%%%%%%%%%%%%%%%%%%
\section{Results}
\label{sec:res}

%\begin{figure}[t]
%\centering
%\includegraphics[width=0.60\textwidth]{figures/spectra_2010_deuteron_EPOS_d_2_final_T=150MeV.png}
%\caption{
%Preliminary transverse momentum spectra in rapidity slices for deuterons produced in inelastic \pp interactions at 158~\GeV. Solid lines show the overlaid two-parameter thermal model with the shape parameter set to $T=150$~MeV (from ~\cite{NA61SHINE:2017fne}) prior to the fit. Only the amplitude parameter was fitted to the data.}
%\label{fig:newpp158results}
%\end{figure}

\begin{figure}[t]
  \centering
  \begin{tabular}{ll}
    \includegraphics[width=0.47\textwidth]{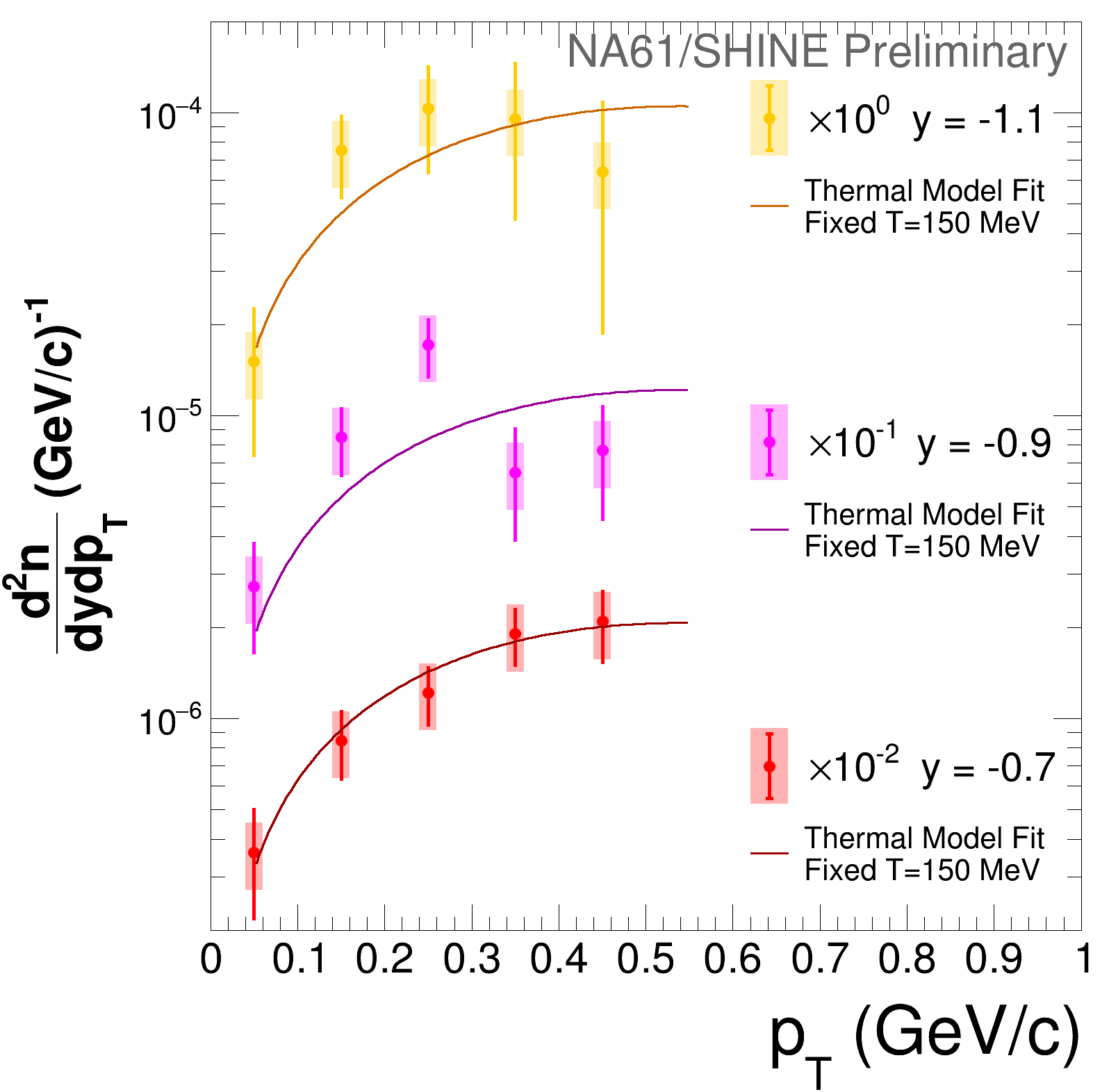} &
    \includegraphics[width=0.47\textwidth]{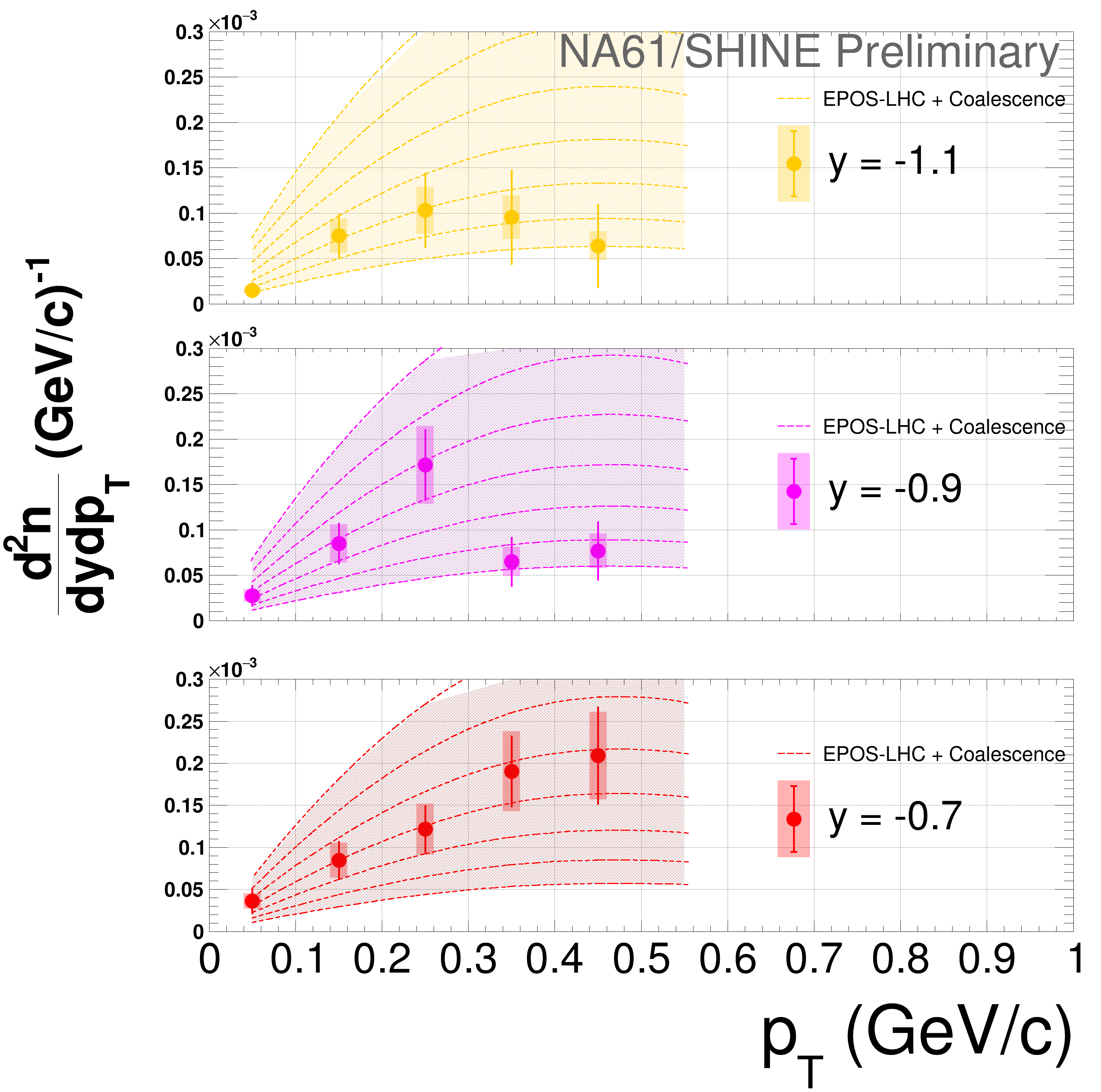}
  \end{tabular}
  \caption{
  Left: Preliminary transverse momentum spectra in rapidity slices for deuterons in inelastic \pp interactions at 158~\GeVc. Solid lines show the overlaid two-parameter thermal model with the shape parameter fixed to $T=150$~MeV (from Ref.~\cite{Aduszkiewicz:2017sei}). Only the amplitude parameter was fitted to the data.
  Right: Preliminary deuteron transverse momentum spectra in rapidity slices compared to the coalescence model predictions from Refs.~\cite{Gomez-Coral:2018yuk, Shukla:2020bql}.
  }
  \label{fig:newpp158results}
\end{figure}

The preliminary transverse momentum spectra in rapidity slices for deuterons produced in inelastic \pp interactions at 158~\GeVc are presented in Fig.~\Ref{fig:newpp158results}.
Figure~\ref{fig:newpp158results} (left) shows an overlay of the thermal model (solid lines) with the data. The thermal model is characterized by two parameters: the temperature $T$, which defines the shape of the distribution, and an amplitude factor. For illustrative purposes, the temperature has been set to $T=150$~MeV, consistent with measurements from Ref.~\cite{Aduszkiewicz:2017sei}), and only the amplitude was fitted to the data. The overlay demonstrates the agreement between the data and the thermal model.
Figure~\Ref{fig:newpp158results} (right) presents a comparison with the predicted spectra band from the coalescence model developed in Refs.~\cite{Gomez-Coral:2018yuk, Shukla:2020bql}. Both the thermal and coalescence models show good agreement with the experimental data within the current level of uncertainties.

Detailed cross-checks have been developed to account for the deuterons produced by interaction of secondary protons inside the detector. The contribution of secondary proton interactions with the right target holder window was estimated to be negligible~\cite{Shukla:2859334}. The contribution of secondary protons reinteractions within the liquid hydrogen of the target was estimated to be less than $\sim$2$\%$~\cite{Shukla:2859334}.
Further studies to estimate the secondary deuteron background using \textsf{Geant4} simulation are ongoing.

%%%%%%%%%%%%%%%%%%%%%%%%%%%%%%%%%%%%%%%%%%%%%%%%%%%%%%%%%%%%%%%%
\section{Summary and Outlook}
\label{sec:con}

Following the success of the first deuteron yield measurement in the \pp 158~\GeVc data set using the $tof$--$\text{d}E/\text{d}x$ analysis presented in this work, the next step was to look at the negatively charged tracks. About $5\times 10^4$ antiprotons were identified by the preliminary application of the $tof-\text{d}E/\text{d}x$ analysis framework developed in the previous sections to analyze negatively charged tracks.
For the same collision system, antideuteron production is about \til1000 times smaller than antiproton production~\cite{Shukla:2020bql}. Applying this to all measurable phase space bins leads to an expectation of about 50 antideuterons in the existing \pp 158~\GeVc data sets. A few such \pp collision events with an antideuteron candidate, which pass all quality and selection cuts, have been identified.

The identification of antideuteron candidates in existing data motivates continued measurements with the upgraded \NASixtyOne detector. 
During the CERN Long Shutdown~2, new TPC backend electronics resulted in a reduction of noise as well as improvement in the $\text{d}E/\text{d}x$ resolution. Upgraded TPC readout electronics increased the data-taking rate to 1.6\,kHz (\til20$\times$ faster). New ToF detector with better timing resolution have also been installed ~\cite{Gazdzicki:2692088, Fields:2739340}.
%Additional TPCs increase the acceptance for highly forward-boosted tracks. 

In October 2025, the CERN SPS accelerator will deliver an unprecedented 600 million \pp collisions at 300~\GeVc to \NASixtyOne, corresponding to a center-of-mass energy of approximately 24~\GeV. Collisions at this energy are particularly relevant for studying the production of secondary antinuclei in cosmic rays. This $10\times$ larger dataset will not only complement existing measurements but also offer a $3\times$ improvement in statistical precision. With the increased energy and number of events, an estimated 3000 deuterons and 100 antideuterons could potentially be identified~\cite{vonDoetinchem:2914265}.
The anticipated reduction in systematic and statistical uncertainties will be instrumental in building and validating new, updated models for the production of astrophysical antideuterons. This could be an essential breakthrough to understand the nature of dark matter.

%%%%%%%%%%%%%%%%%%%%%%%%%%%%%%%%%%%%%%%%%%%%%%%%%%%%%%%%%%%%%%%%
\newpage
\section*{Acknowledgments}
We would like to thank the CERN EP, BE, HSE and EN Departments for the
strong support of \NASixtyOne.
This work was supported by
the Hungarian Scientific Research Fund (grant NKFIH 138136\slash137812\slash138152 and TKP2021-NKTA-64),
the Polish Ministry of Science and Higher Education
(DIR\slash WK\slash\-2016\slash 2017\slash\-10-1, WUT ID-UB), the National Science Centre Poland (grants
2014\slash 14\slash E\slash ST2\slash 00018, %AR, settled
2016\slash 21\slash D\slash ST2\slash 01983, %MMP, settled
2017\slash 25\slash N\slash ST2\slash 02575, %AT, settled
2018\slash 29\slash N\slash ST2\slash 02595, %AM, completed, not settled
2018\slash 30\slash A\slash ST2\slash 00226, %MG, in progress
2018\slash 31\slash G\slash ST2\slash 03910, %SK, in progress
2020\slash 39\slash O\slash ST2\slash 00277), %MR, in progress
the Norwegian Financial Mechanism 2014--2021 (grant 2019\slash 34\slash H\slash ST2\slash 00585),
the Polish Minister of Education and Science (contract No. 2021\slash WK\slash 10),
%the Russian Science Foundation (grant 17-72-20045),
%the Russian Academy of Science and the
%Russian Foundation for Basic Research (grants 08-02-00018, 09-02-00664 and 12-02-91503-CERN),
%the Russian Foundation for Basic Research (RFBR) funding within the research project no. 18-02-40086,
%the Ministry of Science and Higher Education of the Russian Federation, Project "Fundamental properties of elementary particles and cosmology" No 0723-2020-0041,
the European Union's Horizon 2020 research and innovation programme under grant agreement No. 871072,
the Ministry of Education, Culture, Sports,
Science and Tech\-no\-lo\-gy, Japan, Grant-in-Aid for Sci\-en\-ti\-fic
Research (grants 18071005, 19034011, 19740162, 20740160 and 20039012,22H04943),
the German Research Foundation DFG (grants GA\,1480\slash8-1 and project 426579465),
the Bulgarian Ministry of Education and Science within the National
Roadmap for Research Infrastructures 2020--2027, contract No. D01-374/18.12.2020,
Serbian Ministry of Science, Technological Development and Innovation (grant
OI171002), Swiss Nationalfonds Foundation (grant 200020\-117913/1),
ETH Research Grant TH-01\,07-3, National Science Foundation grants
PHY-2013228 and PHY-2411633 and the Fermi National Accelerator Laboratory (Fermilab),
a U.S. Department of Energy, Office of Science, HEP User Facility
managed by Fermi Research Alliance, LLC (FRA), acting under Contract
No. DE-AC02-07CH11359 and the IN2P3-CNRS (France).\\

\newpage
\bibliographystyle{na61Utphys}
{\footnotesize\raggedright
\bibliography{na61_pp158_deuterons}
}

%%%%%%%%%%%%%%%%%%%%%%%%%%%%%%%%%%%%%%%%%%%%%%%%%%%%%%%%%%%%%%%%
%%%%%%%%%%%%% COLLABORATION AUTHORS %%%%%%%%%%%%%%%%%%%%%%%%%%%%
%%%%%%%%%%%%%%%%%%%%%%%%%%%%%%%%%%%%%%%%%%%%%%%%%%%%%%%%%%%%%%%%

\newpage
{\Large The \NASixtyOne Collaboration}
\bigskip
\begin{sloppypar}
% based on XML DB with time Fri Sep 20 11:20:54 2024
% Authors in alphabetical order.

\noindent
{H.\;Adhikary\,\href{https://orcid.org/0000-0002-5746-1268}{\includegraphics[height=1.7ex]{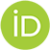}}\textsuperscript{\,11}},
{P.\;Adrich\,\href{https://orcid.org/0000-0002-7019-5451}{\includegraphics[height=1.7ex]{orcid-logo.png}}\textsuperscript{\,13}},
{K.K.\;Allison\,\href{https://orcid.org/0000-0002-3494-9383}{\includegraphics[height=1.7ex]{orcid-logo.png}}\textsuperscript{\,24}},
{N.\;Amin\,\href{https://orcid.org/0009-0004-7572-3817}{\includegraphics[height=1.7ex]{orcid-logo.png}}\textsuperscript{\,4}},
{E.V.\;Andronov\,\href{https://orcid.org/0000-0003-0437-9292}{\includegraphics[height=1.7ex]{orcid-logo.png}}\textsuperscript{\,20}},
{I.-C.\;Arsene\,\href{https://orcid.org/0000-0003-2316-9565}{\includegraphics[height=1.7ex]{orcid-logo.png}}\textsuperscript{\,10}},
{M.\;Bajda\,\href{https://orcid.org/0009-0005-8859-1099}{\includegraphics[height=1.7ex]{orcid-logo.png}}\textsuperscript{\,14}},
{Y.\;Balkova\,\href{https://orcid.org/0000-0002-6957-573X}{\includegraphics[height=1.7ex]{orcid-logo.png}}\textsuperscript{\,16}},
{D.\;Battaglia\,\href{https://orcid.org/0000-0002-5283-0992}{\includegraphics[height=1.7ex]{orcid-logo.png}}\textsuperscript{\,23}},
{A.\;Bazgir\,\href{https://orcid.org/0000-0003-0358-0576}{\includegraphics[height=1.7ex]{orcid-logo.png}}\textsuperscript{\,11}},
{J.\;Bennemann\,\href{https://orcid.org/0009-0007-2300-3799}{\includegraphics[height=1.7ex]{orcid-logo.png}}\textsuperscript{\,4}},
{S.\;Bhosale\,\href{https://orcid.org/0000-0001-5709-4747}{\includegraphics[height=1.7ex]{orcid-logo.png}}\textsuperscript{\,12}},
{M.\;Bielewicz\,\href{https://orcid.org/0000-0001-8267-4874}{\includegraphics[height=1.7ex]{orcid-logo.png}}\textsuperscript{\,13}},
{A.\;Blondel\,\href{https://orcid.org/0000-0002-1597-8859}{\includegraphics[height=1.7ex]{orcid-logo.png}}\textsuperscript{\,3}},
{M.\;Bogomilov\,\href{https://orcid.org/0000-0001-7738-2041}{\includegraphics[height=1.7ex]{orcid-logo.png}}\textsuperscript{\,2}},
{Y.\;Bondar\,\href{https://orcid.org/0000-0003-2773-9668}{\includegraphics[height=1.7ex]{orcid-logo.png}}\textsuperscript{\,11}},
{A.\;Bravar\,\href{https://orcid.org/0000-0002-1134-1527}{\includegraphics[height=1.7ex]{orcid-logo.png}}\textsuperscript{\,27}},
{W.\;Bryli\'nski\,\href{https://orcid.org/0000-0002-3457-6601}{\includegraphics[height=1.7ex]{orcid-logo.png}}\textsuperscript{\,19}},
{J.\;Brzychczyk\,\href{https://orcid.org/0000-0001-5320-6748}{\includegraphics[height=1.7ex]{orcid-logo.png}}\textsuperscript{\,14}},
{M.\;Buryakov\,\href{https://orcid.org/0009-0008-2394-4967}{\includegraphics[height=1.7ex]{orcid-logo.png}}\textsuperscript{\,20}},
{A.F.\;Camino\textsuperscript{\,26}},
{M.\;\'Cirkovi\'c\,\href{https://orcid.org/0000-0002-4420-9688}{\includegraphics[height=1.7ex]{orcid-logo.png}}\textsuperscript{\,21}},
{M.\;Csan\'ad\,\href{https://orcid.org/0000-0002-3154-6925}{\includegraphics[height=1.7ex]{orcid-logo.png}}\textsuperscript{\,6}},
{J.\;Cybowska\,\href{https://orcid.org/0000-0003-2568-3664}{\includegraphics[height=1.7ex]{orcid-logo.png}}\textsuperscript{\,19}},
{T.\;Czopowicz\,\href{https://orcid.org/0000-0003-1908-2977}{\includegraphics[height=1.7ex]{orcid-logo.png}}\textsuperscript{\,11}},
{C.\;Dalmazzone\,\href{https://orcid.org/0000-0001-6945-5845}{\includegraphics[height=1.7ex]{orcid-logo.png}}\textsuperscript{\,3}},
{N.\;Davis\,\href{https://orcid.org/0000-0003-3047-6854}{\includegraphics[height=1.7ex]{orcid-logo.png}}\textsuperscript{\,12}},
{A.\;Dmitriev\,\href{https://orcid.org/0000-0001-7853-0173}{\includegraphics[height=1.7ex]{orcid-logo.png}}\textsuperscript{\,20}},
{P.~von\;Doetinchem\,\href{https://orcid.org/0000-0002-7801-3376}{\includegraphics[height=1.7ex]{orcid-logo.png}}\textsuperscript{\,25}},
{W.\;Dominik\,\href{https://orcid.org/0000-0001-7444-9239}{\includegraphics[height=1.7ex]{orcid-logo.png}}\textsuperscript{\,17}},
{J.\;Dumarchez\,\href{https://orcid.org/0000-0002-9243-4425}{\includegraphics[height=1.7ex]{orcid-logo.png}}\textsuperscript{\,3}},
{R.\;Engel\,\href{https://orcid.org/0000-0003-2924-8889}{\includegraphics[height=1.7ex]{orcid-logo.png}}\textsuperscript{\,4}},
{G.A.\;Feofilov\,\href{https://orcid.org/0000-0003-3700-8623}{\includegraphics[height=1.7ex]{orcid-logo.png}}\textsuperscript{\,20}},
{L.\;Fields\,\href{https://orcid.org/0000-0001-8281-3686}{\includegraphics[height=1.7ex]{orcid-logo.png}}\textsuperscript{\,23}},
{Z.\;Fodor\,\href{https://orcid.org/0000-0003-2519-5687}{\includegraphics[height=1.7ex]{orcid-logo.png}}\textsuperscript{\,5,18}},
{M.\;Friend\,\href{https://orcid.org/0000-0003-4660-4670}{\includegraphics[height=1.7ex]{orcid-logo.png}}\textsuperscript{\,7}},
{M.\;Ga\'zdzicki\,\href{https://orcid.org/0000-0002-6114-8223}{\includegraphics[height=1.7ex]{orcid-logo.png}}\textsuperscript{\,11}},
{K.E.\;Gollwitzer\textsuperscript{\,22}},
{O.\;Golosov\,\href{https://orcid.org/0000-0001-6562-2925}{\includegraphics[height=1.7ex]{orcid-logo.png}}\textsuperscript{\,20}},
{V.\;Golovatyuk\,\href{https://orcid.org/0009-0006-5201-0990}{\includegraphics[height=1.7ex]{orcid-logo.png}}\textsuperscript{\,20}},
{M.\;Golubeva\,\href{https://orcid.org/0009-0003-4756-2449}{\includegraphics[height=1.7ex]{orcid-logo.png}}\textsuperscript{\,20}},
{K.\;Grebieszkow\,\href{https://orcid.org/0000-0002-6754-9554}{\includegraphics[height=1.7ex]{orcid-logo.png}}\textsuperscript{\,19}},
{F.\;Guber\,\href{https://orcid.org/0000-0001-8790-3218}{\includegraphics[height=1.7ex]{orcid-logo.png}}\textsuperscript{\,20}},
{P.G.\;Hurh\,\href{https://orcid.org/0000-0002-9024-5399}{\includegraphics[height=1.7ex]{orcid-logo.png}}\textsuperscript{\,22}},
{S.\;Ilieva\,\href{https://orcid.org/0000-0001-9204-2563}{\includegraphics[height=1.7ex]{orcid-logo.png}}\textsuperscript{\,2}},
{A.\;Ivashkin\,\href{https://orcid.org/0000-0003-4595-5866}{\includegraphics[height=1.7ex]{orcid-logo.png}}\textsuperscript{\,20}},
{A.\;Izvestnyy\,\href{https://orcid.org/0009-0009-1305-7309}{\includegraphics[height=1.7ex]{orcid-logo.png}}\textsuperscript{\,20}},
{N.\;Karpushkin\,\href{https://orcid.org/0000-0001-5513-9331}{\includegraphics[height=1.7ex]{orcid-logo.png}}\textsuperscript{\,20}},
{M.\;Kie{\l}bowicz\,\href{https://orcid.org/0000-0002-4403-9201}{\includegraphics[height=1.7ex]{orcid-logo.png}}\textsuperscript{\,12}},
{V.A.\;Kireyeu\,\href{https://orcid.org/0000-0002-5630-9264}{\includegraphics[height=1.7ex]{orcid-logo.png}}\textsuperscript{\,20}},
{R.\;Kolesnikov\,\href{https://orcid.org/0009-0006-4224-1058}{\includegraphics[height=1.7ex]{orcid-logo.png}}\textsuperscript{\,20}},
{D.\;Kolev\,\href{https://orcid.org/0000-0002-9203-4739}{\includegraphics[height=1.7ex]{orcid-logo.png}}\textsuperscript{\,2}},
{Y.\;Koshio\,\href{https://orcid.org/0000-0003-0437-8505}{\includegraphics[height=1.7ex]{orcid-logo.png}}\textsuperscript{\,8}},
{S.\;Kowalski\,\href{https://orcid.org/0000-0001-9888-4008}{\includegraphics[height=1.7ex]{orcid-logo.png}}\textsuperscript{\,16}},
{B.\;Koz{\l}owski\,\href{https://orcid.org/0000-0001-8442-2320}{\includegraphics[height=1.7ex]{orcid-logo.png}}\textsuperscript{\,19}},
{A.\;Krasnoperov\,\href{https://orcid.org/0000-0002-1425-2861}{\includegraphics[height=1.7ex]{orcid-logo.png}}\textsuperscript{\,20}},
{W.\;Kucewicz\,\href{https://orcid.org/0000-0002-2073-711X}{\includegraphics[height=1.7ex]{orcid-logo.png}}\textsuperscript{\,15}},
{M.\;Kuchowicz\,\href{https://orcid.org/0000-0003-3174-585X}{\includegraphics[height=1.7ex]{orcid-logo.png}}\textsuperscript{\,18}},
{M.\;Kuich\,\href{https://orcid.org/0000-0002-6507-8699}{\includegraphics[height=1.7ex]{orcid-logo.png}}\textsuperscript{\,17}},
{A.\;Kurepin\,\href{https://orcid.org/0000-0002-1851-4136}{\includegraphics[height=1.7ex]{orcid-logo.png}}\textsuperscript{\,20}},
{A.\;L\'aszl\'o\,\href{https://orcid.org/0000-0003-2712-6968}{\includegraphics[height=1.7ex]{orcid-logo.png}}\textsuperscript{\,5}},
{M.\;Lewicki\,\href{https://orcid.org/0000-0002-8972-3066}{\includegraphics[height=1.7ex]{orcid-logo.png}}\textsuperscript{\,12}},
{G.\;Lykasov\,\href{https://orcid.org/0000-0002-1544-6959}{\includegraphics[height=1.7ex]{orcid-logo.png}}\textsuperscript{\,20}},
{J.\;Lyon\,\href{https://orcid.org/0009-0003-2579-8821}{\includegraphics[height=1.7ex]{orcid-logo.png}}\textsuperscript{\,25}},
{V.V.\;Lyubushkin\,\href{https://orcid.org/0000-0003-0136-233X}{\includegraphics[height=1.7ex]{orcid-logo.png}}\textsuperscript{\,20}},
{M.\;Ma\'ckowiak-Paw{\l}owska\,\href{https://orcid.org/0000-0003-3954-6329}{\includegraphics[height=1.7ex]{orcid-logo.png}}\textsuperscript{\,19}},
{A.\;Makhnev\,\href{https://orcid.org/0009-0002-9745-1897}{\includegraphics[height=1.7ex]{orcid-logo.png}}\textsuperscript{\,20}},
{B.\;Maksiak\,\href{https://orcid.org/0000-0002-7950-2307}{\includegraphics[height=1.7ex]{orcid-logo.png}}\textsuperscript{\,13}},
{A.I.\;Malakhov\,\href{https://orcid.org/0000-0001-8569-8409}{\includegraphics[height=1.7ex]{orcid-logo.png}}\textsuperscript{\,20}},
{A.\;Marcinek\,\href{https://orcid.org/0000-0001-9922-743X}{\includegraphics[height=1.7ex]{orcid-logo.png}}\textsuperscript{\,12}},
{A.D.\;Marino\,\href{https://orcid.org/0000-0002-1709-538X}{\includegraphics[height=1.7ex]{orcid-logo.png}}\textsuperscript{\,24}},
{H.-J.\;Mathes\,\href{https://orcid.org/0000-0002-0680-040X}{\includegraphics[height=1.7ex]{orcid-logo.png}}\textsuperscript{\,4}},
{T.\;Matulewicz\,\href{https://orcid.org/0000-0003-2098-1216}{\includegraphics[height=1.7ex]{orcid-logo.png}}\textsuperscript{\,17}},
{V.\;Matveev\,\href{https://orcid.org/0000-0002-2745-5908}{\includegraphics[height=1.7ex]{orcid-logo.png}}\textsuperscript{\,20}},
{G.L.\;Melkumov\,\href{https://orcid.org/0009-0004-2074-6755}{\includegraphics[height=1.7ex]{orcid-logo.png}}\textsuperscript{\,20}},
{A.\;Merzlaya\,\href{https://orcid.org/0000-0002-6553-2783}{\includegraphics[height=1.7ex]{orcid-logo.png}}\textsuperscript{\,10}},
{{\L}.\;Mik\,\href{https://orcid.org/0000-0003-2712-6861}{\includegraphics[height=1.7ex]{orcid-logo.png}}\textsuperscript{\,15}},
{S.\;Morozov\,\href{https://orcid.org/0000-0002-6748-7277}{\includegraphics[height=1.7ex]{orcid-logo.png}}\textsuperscript{\,20}},
{Y.\;Nagai\,\href{https://orcid.org/0000-0002-1792-5005}{\includegraphics[height=1.7ex]{orcid-logo.png}}\textsuperscript{\,6}},
{T.\;Nakadaira\,\href{https://orcid.org/0000-0003-4327-7598}{\includegraphics[height=1.7ex]{orcid-logo.png}}\textsuperscript{\,7}},
{M.\;Naskr\k{e}t\,\href{https://orcid.org/0000-0002-5634-6639}{\includegraphics[height=1.7ex]{orcid-logo.png}}\textsuperscript{\,18}},
{S.\;Nishimori\,\href{https://orcid.org/~0000-0002-1820-0938}{\includegraphics[height=1.7ex]{orcid-logo.png}}\textsuperscript{\,7}},
{A.\;Olivier\,\href{https://orcid.org/0000-0003-4261-8303}{\includegraphics[height=1.7ex]{orcid-logo.png}}\textsuperscript{\,23}},
{V.\;Ozvenchuk\,\href{https://orcid.org/0000-0002-7821-7109}{\includegraphics[height=1.7ex]{orcid-logo.png}}\textsuperscript{\,12}},
{O.\;Panova\,\href{https://orcid.org/0000-0001-5039-7788}{\includegraphics[height=1.7ex]{orcid-logo.png}}\textsuperscript{\,11}},
{V.\;Paolone\,\href{https://orcid.org/0000-0003-2162-0957}{\includegraphics[height=1.7ex]{orcid-logo.png}}\textsuperscript{\,26}},
{O.\;Petukhov\,\href{https://orcid.org/0000-0002-8872-8324}{\includegraphics[height=1.7ex]{orcid-logo.png}}\textsuperscript{\,20}},
{I.\;Pidhurskyi\,\href{https://orcid.org/0000-0001-9916-9436}{\includegraphics[height=1.7ex]{orcid-logo.png}}\textsuperscript{\,11}},
{R.\;P{\l}aneta\,\href{https://orcid.org/0000-0001-8007-8577}{\includegraphics[height=1.7ex]{orcid-logo.png}}\textsuperscript{\,14}},
{P.\;Podlaski\,\href{https://orcid.org/0000-0002-0232-9841}{\includegraphics[height=1.7ex]{orcid-logo.png}}\textsuperscript{\,17}},
{B.A.\;Popov\,\href{https://orcid.org/0000-0001-5416-9301}{\includegraphics[height=1.7ex]{orcid-logo.png}}\textsuperscript{\,20,3}},
{B.\;P\'orfy\,\href{https://orcid.org/0000-0001-5724-9737}{\includegraphics[height=1.7ex]{orcid-logo.png}}\textsuperscript{\,5,6}},
{D.S.\;Prokhorova\,\href{https://orcid.org/0000-0003-3726-9196}{\includegraphics[height=1.7ex]{orcid-logo.png}}\textsuperscript{\,20}},
{D.\;Pszczel\,\href{https://orcid.org/0000-0002-4697-6688}{\includegraphics[height=1.7ex]{orcid-logo.png}}\textsuperscript{\,13}},
{S.\;Pu{\l}awski\,\href{https://orcid.org/0000-0003-1982-2787}{\includegraphics[height=1.7ex]{orcid-logo.png}}\textsuperscript{\,16}},
{R.\;Renfordt\,\href{https://orcid.org/0000-0002-5633-104X}{\includegraphics[height=1.7ex]{orcid-logo.png}}\textsuperscript{\,16}},
{L.\;Ren\,\href{https://orcid.org/0000-0003-1709-7673}{\includegraphics[height=1.7ex]{orcid-logo.png}}\textsuperscript{\,24}},
{V.Z.\;Reyna~Ortiz\,\href{https://orcid.org/0000-0002-7026-8198}{\includegraphics[height=1.7ex]{orcid-logo.png}}\textsuperscript{\,11}},
{D.\;R\"ohrich\textsuperscript{\,9}},
{E.\;Rondio\,\href{https://orcid.org/0000-0002-2607-4820}{\includegraphics[height=1.7ex]{orcid-logo.png}}\textsuperscript{\,13}},
{M.\;Roth\,\href{https://orcid.org/0000-0003-1281-4477}{\includegraphics[height=1.7ex]{orcid-logo.png}}\textsuperscript{\,4}},
{{\L}.\;Rozp{\l}ochowski\,\href{https://orcid.org/0000-0003-3680-6738}{\includegraphics[height=1.7ex]{orcid-logo.png}}\textsuperscript{\,12}},
{B.T.\;Rumberger\,\href{https://orcid.org/0000-0002-4867-945X}{\includegraphics[height=1.7ex]{orcid-logo.png}}\textsuperscript{\,24}},
{M.\;Rumyantsev\,\href{https://orcid.org/0000-0001-8233-2030}{\includegraphics[height=1.7ex]{orcid-logo.png}}\textsuperscript{\,20}},
{A.\;Rustamov\,\href{https://orcid.org/0000-0001-8678-6400}{\includegraphics[height=1.7ex]{orcid-logo.png}}\textsuperscript{\,1}},
{M.\;Rybczynski\,\href{https://orcid.org/0000-0002-3638-3766}{\includegraphics[height=1.7ex]{orcid-logo.png}}\textsuperscript{\,11}},
{A.\;Rybicki\,\href{https://orcid.org/0000-0003-3076-0505}{\includegraphics[height=1.7ex]{orcid-logo.png}}\textsuperscript{\,12}},
{D.\;Rybka\,\href{https://orcid.org/0000-0002-9924-6398}{\includegraphics[height=1.7ex]{orcid-logo.png}}\textsuperscript{\,13}},
{K.\;Sakashita\,\href{https://orcid.org/0000-0003-2602-7837}{\includegraphics[height=1.7ex]{orcid-logo.png}}\textsuperscript{\,7}},
{K.\;Schmidt\,\href{https://orcid.org/0000-0002-0903-5790}{\includegraphics[height=1.7ex]{orcid-logo.png}}\textsuperscript{\,16}},
{A.\;Seryakov\,\href{https://orcid.org/0000-0002-5759-5485}{\includegraphics[height=1.7ex]{orcid-logo.png}}\textsuperscript{\,20}},
{P.\;Seyboth\,\href{https://orcid.org/0000-0002-4821-6105}{\includegraphics[height=1.7ex]{orcid-logo.png}}\textsuperscript{\,11}},
{U.A.\;Shah\,\href{https://orcid.org/0000-0002-9315-1304}{\includegraphics[height=1.7ex]{orcid-logo.png}}\textsuperscript{\,11}},
{Y.\;Shiraishi\,\href{https://orcid.org/0000-0002-0132-3923}{\includegraphics[height=1.7ex]{orcid-logo.png}}\textsuperscript{\,8}},
{A.\;Shukla\,\href{https://orcid.org/0000-0003-3839-7229}{\includegraphics[height=1.7ex]{orcid-logo.png}}\textsuperscript{\,25}},
{M.\;S{\l}odkowski\,\href{https://orcid.org/0000-0003-0463-2753}{\includegraphics[height=1.7ex]{orcid-logo.png}}\textsuperscript{\,19}},
{P.\;Staszel\,\href{https://orcid.org/0000-0003-4002-1626}{\includegraphics[height=1.7ex]{orcid-logo.png}}\textsuperscript{\,14}},
{G.\;Stefanek\,\href{https://orcid.org/0000-0001-6656-9177}{\includegraphics[height=1.7ex]{orcid-logo.png}}\textsuperscript{\,11}},
{J.\;Stepaniak\,\href{https://orcid.org/0000-0003-2064-9870}{\includegraphics[height=1.7ex]{orcid-logo.png}}\textsuperscript{\,13}},
{F.\;Sutter\textsuperscript{\,4}},
{{\L}.\;\'Swiderski\,\href{https://orcid.org/0000-0001-5857-2085}{\includegraphics[height=1.7ex]{orcid-logo.png}}\textsuperscript{\,13}},
{J.\;Szewi\'nski\,\href{https://orcid.org/0000-0003-2981-9303}{\includegraphics[height=1.7ex]{orcid-logo.png}}\textsuperscript{\,13}},
{R.\;Szukiewicz\,\href{https://orcid.org/0000-0002-1291-4040}{\includegraphics[height=1.7ex]{orcid-logo.png}}\textsuperscript{\,18}},
{A.\;Taranenko\,\href{https://orcid.org/0000-0003-1737-4474}{\includegraphics[height=1.7ex]{orcid-logo.png}}\textsuperscript{\,20}},
{A.\;Tefelska\,\href{https://orcid.org/0000-0002-6069-4273}{\includegraphics[height=1.7ex]{orcid-logo.png}}\textsuperscript{\,19}},
{D.\;Tefelski\,\href{https://orcid.org/0000-0003-0802-2290}{\includegraphics[height=1.7ex]{orcid-logo.png}}\textsuperscript{\,19}},
{V.\;Tereshchenko\textsuperscript{\,20}},
{R.\;Tsenov\,\href{https://orcid.org/0000-0002-1330-8640}{\includegraphics[height=1.7ex]{orcid-logo.png}}\textsuperscript{\,2}},
{L.\;Turko\,\href{https://orcid.org/0000-0002-5474-8650}{\includegraphics[height=1.7ex]{orcid-logo.png}}\textsuperscript{\,18}},
{T.S.\;Tveter\,\href{https://orcid.org/0009-0003-7140-8644}{\includegraphics[height=1.7ex]{orcid-logo.png}}\textsuperscript{\,10}},
{M.\;Unger\,\href{https://orcid.org/0000-0002-7651-0272~}{\includegraphics[height=1.7ex]{orcid-logo.png}}\textsuperscript{\,4}},
{M.\;Urbaniak\,\href{https://orcid.org/0000-0002-9768-030X}{\includegraphics[height=1.7ex]{orcid-logo.png}}\textsuperscript{\,16}},
{D.\;Veberi\v{c}\,\href{https://orcid.org/0000-0003-2683-1526}{\includegraphics[height=1.7ex]{orcid-logo.png}}\textsuperscript{\,4}},
{O.\;Vitiuk\,\href{https://orcid.org/0000-0002-9744-3937}{\includegraphics[height=1.7ex]{orcid-logo.png}}\textsuperscript{\,18}},
{V.\;Volkov\,\href{https://orcid.org/0000-0002-4785-7517}{\includegraphics[height=1.7ex]{orcid-logo.png}}\textsuperscript{\,20}},
{A.\;Wickremasinghe\,\href{https://orcid.org/0000-0002-5325-0455}{\includegraphics[height=1.7ex]{orcid-logo.png}}\textsuperscript{\,22}},
{K.\;Witek\,\href{https://orcid.org/0009-0004-6699-1895}{\includegraphics[height=1.7ex]{orcid-logo.png}}\textsuperscript{\,15}},
{K.\;W\'ojcik\,\href{https://orcid.org/0000-0002-8315-9281}{\includegraphics[height=1.7ex]{orcid-logo.png}}\textsuperscript{\,16}},
{O.\;Wyszy\'nski\,\href{https://orcid.org/0000-0002-6652-0450}{\includegraphics[height=1.7ex]{orcid-logo.png}}\textsuperscript{\,11}},
{A.\;Zaitsev\,\href{https://orcid.org/0000-0003-4711-9925}{\includegraphics[height=1.7ex]{orcid-logo.png}}\textsuperscript{\,20}},
{E.\;Zherebtsova\,\href{https://orcid.org/0000-0002-1364-0969}{\includegraphics[height=1.7ex]{orcid-logo.png}}\textsuperscript{\,18}},
{E.D.\;Zimmerman\,\href{https://orcid.org/0000-0002-6394-6659}{\includegraphics[height=1.7ex]{orcid-logo.png}}\textsuperscript{\,24}}, and
{A.\;Zviagina\,\href{https://orcid.org/0009-0007-5211-6493}{\includegraphics[height=1.7ex]{orcid-logo.png}}\textsuperscript{\,20}}

\end{sloppypar}
% based on XML DB with time Fri Sep 20 11:20:54 2024
% Institutes in alphabetical order.

\noindent
\textsuperscript{1}~National Nuclear Research Center, Baku, Azerbaijan\\
\textsuperscript{2}~Faculty of Physics, University of Sofia, Sofia, Bulgaria\\
\textsuperscript{3}~LPNHE, Sorbonne University, CNRS/IN2P3, Paris, France\\
\textsuperscript{4}~Karlsruhe Institute of Technology, Karlsruhe, Germany\\
\textsuperscript{5}~HUN-REN Wigner Research Centre for Physics, Budapest, Hungary\\
\textsuperscript{6}~E\"otv\"os Lor\'and University, Budapest, Hungary\\
\textsuperscript{7}~Institute for Particle and Nuclear Studies, Tsukuba, Japan\\
\textsuperscript{8}~Okayama University, Japan\\
\textsuperscript{9}~University of Bergen, Bergen, Norway\\
\textsuperscript{10}~University of Oslo, Oslo, Norway\\
\textsuperscript{11}~Jan Kochanowski University, Kielce, Poland\\
\textsuperscript{12}~Institute of Nuclear Physics, Polish Academy of Sciences, Cracow, Poland\\
\textsuperscript{13}~National Centre for Nuclear Research, Warsaw, Poland\\
\textsuperscript{14}~Jagiellonian University, Cracow, Poland\\
\textsuperscript{15}~AGH - University of Krakow, Poland\\
\textsuperscript{16}~University of Silesia, Katowice, Poland\\
\textsuperscript{17}~University of Warsaw, Warsaw, Poland\\
\textsuperscript{18}~University of Wroc{\l}aw,  Wroc{\l}aw, Poland\\
\textsuperscript{19}~Warsaw University of Technology, Warsaw, Poland\\
\textsuperscript{20}~Affiliated with an institution covered by a cooperation agreement with CERN\\
\textsuperscript{21}~University of Belgrade, Belgrade, Serbia\\
\textsuperscript{22}~Fermilab, Batavia, USA\\
\textsuperscript{23}~University of Notre Dame, Notre Dame, USA\\
\textsuperscript{24}~University of Colorado, Boulder, USA\\
\textsuperscript{25}~University of Hawaii at Manoa, Honolulu, USA\\
\textsuperscript{26}~University of Pittsburgh, Pittsburgh, USA\\
\textsuperscript{27}~University of Geneva, Geneva, Switzerland\footnote{No longer affiliated with the NA61/SHINE collaboration}\\

%%%%%%%%%%%%%%%%%%%%%%%%%%%%%%%%%%%%%%%%%%%%%%%%%%%%%%%%%%%%%%%%

\end{document}